\documentclass[12pt]{iopart}

\usepackage{iopams}
\usepackage{graphicx}
\usepackage{bm}
\usepackage{citesort}

\begin{document}

\title[Sample]{Uniform magnetization dynamics of a submicron ferromagnetic coin driven by the spin-orbit coupled spin torque}
\author{P. V. Bondarenko$^{1,2,a)}$, E. Ya. Sherman$^{1,3,b)}$}
\ead{$^{a)}$paulvbond@gmail.com, $^{b)}$evgeny.sherman@ehu.es}
\address{$^1$Department of Physical Chemistry, Universidad del Pa\'{\i}s Vasco UPV-EHU, 48080, Bilbao, Spain}
\address{$^2$Institute of Magnetism, National Academy of Sciences, 03142 Kyiv, Ukraine}
\address{$^3$IKERBASQUE Basque Foundation for Science, Bilbao, Spain}
\date{\today}

\begin{abstract}
A simple model of magnetization dynamics in a ferromagnet/doped semiconductor hybrid structure with Rashba spin-orbit interaction (SOI) driven by an applied pulse of the electric field is proposed.
The electric current excited by the applied field is spin-polarized due to the SOI and therefore it induces the magnetization rotation in the ferromagnetic layer via s-d exchange coupling.
Magnetization dynamics dependence on the electric pulse shape and magnitude is analyzed for realistic values of parameters.
We show that it is similar to the dynamics of a damped nonlinear oscillator with the time-dependent frequency proportional to the square root of the applied electric field.
The magnetization switching properties of an elliptic magnetic element are examined as a function of the applied field magnitude and direction.
\end{abstract}
\pacs{75.75.-c, 75.78.-n, 85.75.-d}
\vspace{2pc}
\noindent{\it Keywords}: {Rashba spin-orbit interaction, macrospin, magnetization switching, spin-orbit torque}

\maketitle

Modern logic devices are based on driven charge dynamics in semiconductor structures, whereas modern data-storage devices utilize magnetization switching in multilayers of magnetic metals and insulators \cite{awschalom2007,nowack2007}.
New hybrid spintronics devices based on ferromagnet/paramagnet or ferromagnet/doped semiconductor structures could unite these tasks by performing logic operations and data storage.
Charge current and magnetization in these heterostructures are related through the spin-orbit (SOI) and s-d exchange interactions \cite{das_sarma2003,papp2005,jungwirth2006,zutic2004}.

The spin-orbit interaction of a moving electron spin with the electric field appears due to the existence of the magnetic field in the electron rest frame.
This electric field arises from the inversion asymmetry of conducting structure, which leads to the effective Bychkov-Rashba SOI \cite{bychkov1984,winkler2003}.
Therefore electron current is spin-polarized transversely to its direction due to the SOI as was proposed by Aronov and
Lyanda-Geller \cite{aronov1989} and Edelstein \cite{edelstein1990}.

\begin{figure}[ht]
    \center
    \includegraphics[width=8 cm]{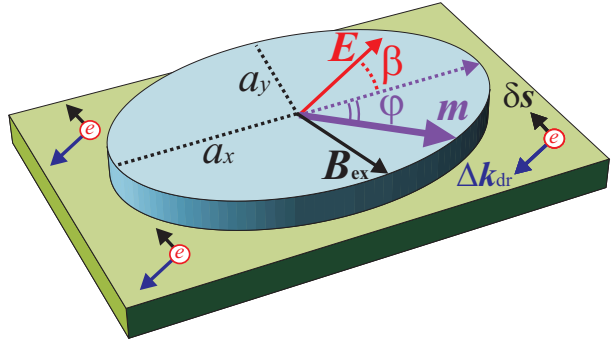}
    \caption{Sketch of the hybrid structure. Magnetic dynamics is induced by the charge current via s-d exchange coupling. Magnetization $\bi{m}$ rotates under the action of  effective exchange magnetic field $\bi{B_{\mathrm{ex}}}(\bi{\delta s})$ at the angle $\varphi$ to the initial direction $\bi{m}_0$.}\label{Schema}
\end{figure}

We consider in-plane magnetization dynamics of a ferromagnetic coin deposited on the surface of a two-dimensional electron gas (2DEG) driven by the charge current as shown in the figure~\ref{Schema}.
Supposed here uniform in-plane magnetization is the steady state for thin (thickness is no larger than the exchange length) and planar (diameter is at least one order of magnitude larger than the exchange length) films \cite{ha2003,metlov2008}.

Current-induced carrier spin density exerts the rotation of the magnetization in the magnetic elements of this heterostructure via s-d exchange coupling, producing the spin-transfer torque effect \cite{ralph2008,haney2013,gambardella2011}.
Therefore one can change the direction of the film magnetization by applying the electric field to the 2DEG substrate \cite{kurebayashi2014,manchon2008,manchon2009,fukami2016,ciccarelli2016,avci2015,garello2013,miron2011}.

The objective of this paper is to investigate magnetization dynamics and its switching modes under the action of applied electric field pulses for realistic values of physical parameters within a simple spin-transfer torque model.
Current-induced dynamics of magnetization reveals a nontrivial dependence on the exchange coupling as has been shown by Yokoyama \cite{yokoyama2011} for a ferromagnet on the surface of a topological insulator.

The effective Hamiltonian for electron in the planar magnetic field with the Bychkov-Rashba SOI \cite{eldridge2008,inoue2003,wang2015} presented as:
\begin{equation}
\mathcal{H}=\frac{\hbar ^{2}k^{2}}{2m^{\ast }}+\frac{\hbar \omega _{m}}{2}\bi{m\cdot}\bm{\hat{\sigma}}+\frac{\hbar \alpha}{2}\left[ \bi{k}\times\bm{\hat{\sigma}}\right]\cdot\bi{e}_{z},
\end{equation}
where $\bi{m}=\bi{M}/M_{0}$ is the unit vector of the in-plane magnetization of the magnetic element, $m^{\ast}$ is the effective electron mass and $\bi{e}_{z}$ is the unit vector normal to 2DEG system.
The value of the s-d exchange energy per carrier $\omega _{m}\approx 2\pi \times 1.6$~THz can be estimated from the values of the kinetic-exchange coupling $J_{\mathrm{ex}}$ between charge carriers and local magnetic moments (see~\eref{energy}) \cite{kurebayashi2014}.
The Rashba coupling constant is $\alpha \approx 10^{7}$~cm/s as can be achieved in the structures based on the InSb \cite{winkler2003,meier2007}.
Thus the effective magnetic field, due to the exchange and SOI terms is
\begin{equation}
\bi{B}_{\mathrm{eff}}=-\frac{1}{\mu_B}\frac{\delta\mathcal{H}}{\delta\bm{\sigma}}=-\frac{\hbar\omega_{m}}{2\mu_{B}}
\bi{m}-\frac{\hbar \alpha }{2\mu _{B}}\left[\bi{e}_{z}\times\bi{k}\right] .  \label{beff}
\end{equation}

As the result of applying an in-plane electric field $\bi{E}$, the electron ensemble receives drift increment of quasi-momentum ${\Delta\bi{k}_{\mathrm{dr}}\approx e\bi{E}\tau _{p}/\hbar}$ where $e$ is the electron charge.
Ensemble momentum relaxation time ${\tau _{p}\sim 10^{-4}}$~ns is obtained from the experimental electron mobility at room temperature \cite{leyland2007}.
For the linear regime, when ${\Delta k_{\mathrm{dr}} \ll k_{T}}$, where $k_{T}$ is the electron thermal wavevector (i.e. $E \ll \hbar k_T/|e|\tau_p\approx10$~kV/cm) we can make an assumption that spin density matrix does not change the shape and just shifts:
$\bm{\hat{\rho}}_{0}\left( \bi{k},t\right) \rightarrow \bm{\hat{\rho}}_{0}\left( \bi{k}+\Delta\bi{k}_{\mathrm{dr}},t\right)$.
As the result, in the first order approximation for $\Delta \bi{k_{\mathrm{dr}}}$, the magnetic susceptibility $\chi _{s}^{(0)}$ of the 2DEG does not change.
Mean spin per electron in the applied electric field taking into account \eref{beff} is:
\begin{equation}
\bi{s}\left( \bi{m},\bi{E}\right) \approx \chi _{s}^{\left(
0\right) }\bi{B}_{\mathrm{eff}}=-s_{0}\left( \bi{m}+\frac{\left[ \bi{E}\times\bi{e}_{z}\right]}{E_0}\right), \label{el_spin}
\end{equation}
where $E_{0}=\hbar \omega _{m}/\alpha |e|\tau _{p}$ and $s_{0}=\chi
_{s}^{\left( 0\right) }\hbar \omega _{m}/2\mu _{B}$. At the room temperature
$E_{0}\approx 7$ kV/cm so ${E/E_0\ll1}$ (considering the thermodynamic limitation of model applicability ${E\ll10}$~kV/cm mentioned above).

The mean spin value $s_{0}\approx 0.05$ could be estimated numerically assuming that $\bm{\hat{\rho}}_{0}\left( \bi{r}%
,\bi{k},t\right) $ is a stationary diagonal operator with elements
representing spin up and spin down ($\sigma=\pm1$) Fermi distribution functions:
\begin{equation*}
\bm{\hat{\rho}}_{0,\sigma \sigma }\left( \bi{k},t\right) =\frac{1}{1+\exp\left[ \left( \varepsilon _{\sigma }\left( \bi{k}\right) -\mu \right)/T\right]},
\end{equation*}
where $\varepsilon _{\sigma }(\bi{k}) =\hbar^2 k^{2}\bi{/}2m^{\ast }+\sigma \mu
_{B} B_{\mathrm{eff}}(\bi{k})$ \cite{fert1969}.
If the effective magnetic field is changed, the electron spin will relax to the
value \eref{el_spin} during characteristic time $\tau_{s}\sim 10^{-2}$~ns which was estimated with the Elliott-Yafet mechanism \cite{zutic2004,gantmakher1987,fishman1977,song2002,tackeuchi1997,khaetskii2000}.
Therefore the spin adjusting time into new position \eref{el_spin} under the action of changing magnetic field can be neglected since the spin relaxation time is considerably smaller than minimal characteristic time of the magnetization dynamics $\sim0.1$~ns as will be shown below.

Magnetic energy density of a thin ferromagnetic film in the case of uniform in-plane magnetization is:%
\begin{equation}
W(\bi{m})=\frac{J_{\mathrm{ex}}n_{\mathrm{el}}M_{0}}{2\mu _{B}}\bi{m}\cdot\bi{s}+2\pi M_{0}^{2}\bi{m}\cdot\left(\bm{\hat{N}}\bi{m}\right).  \label{energy}
\end{equation}
The first term in \eref{energy} describes the s-d exchange interaction,
where we use the following numerical values: $J_{\mathrm{ex}}=55$ meV$\cdot $nm$^{3}$ is the kinetic-exchange coupling constant \cite{kurebayashi2014}, $M_{0}=800$~emu/cm$^3$ - permalloy saturation magnetization and $n_{\mathrm{el}}=5\cdot 10^{17}$ cm$^{-3}$ - electron concentration.
The second term is the dipolar shape anisotropy energy density.
The demagnetization tensor $\bm{\hat{N}}$ is averaged over ferromagnetic particle volume $V_{\rm d}$.
It is symmetrical and has the trace $\Tr(\bm{\hat{N}})=1$ \cite{akhiezer1968}.
The symmetry axes of FM particle are the principal axes of the demagnetization tensor, i.e. it is diagonal, $\bm{\hat{N}}_{ij}=\delta_{ij}N_{i}$, where:
\begin{equation}
N_{i}=-\frac{1}{4\pi V_{\rm d}}\int\limits_{V_d}d^3\bi{r}\frac{\partial^2}{\partial r_{i}^{2}}\int\limits_{V_d}\frac{d^3\bi{r'}}{|\bi{r}-\bi{r'}|}.  \label{demten}
\end{equation}
Let us suppose that ferromagnetic particle has the elliptic cylinder shape with major semi-axis $a_x$, minor semi-axis $a_y$ and height $h$.
Then one can use well-known expressions for the ellipsoid demagnetization tensor coefficients \cite{akhiezer1968}:
\begin{equation}
N_{i}=\frac{a_xa_yh}{2}\int\limits_{0}^{\infty}\frac{ds}{(s+a_{i}^{2})\sqrt{(s+a_{x}^{2})(s+a_{y}^{2})(s+h^{2})}}.
\label{elten}
\end{equation}

Assuming that the uniformly magnetized element is thin ${(h\approx10^{-2}a)}$, the tensor's in-plane components $N_{x,y}\propto h/a$ are small compared to the perpendicular component $N_z\approx1$.
Given that $|\bi{m}|=1$ and substituting \eref{el_spin} into \eref{energy}, the magnetic free energy density \eref{energy} is:
\begin{equation}
    W(\bi{m})\approx\frac{J_{\mathrm{ex}}n_{\mathrm{el}}M_{0}s_0}{2\mu_{B}E_0}\left[\bi{E}\times\bi{m}\right]_z
    +2\pi M_{0}^{2}\left(Km_{y}^2+m_{z}^2\right).
    \label{energyE}
\end{equation}
One can obtain the coefficient $K$ in \eref{energyE} by substitution of the formula \eref{elten} in the anisotropy energy expression:
\begin{equation*}
K=N_y-N_x\approx\frac{h}{a_x}\eta^2\sqrt{1-\eta^2}\left(\frac{3\pi}{16}+\frac{9\pi}{64}\eta^2\right),
\end{equation*}
where $\eta$ is the eccentricity of the base of the ferromagnetic coin, $\eta^2\equiv1-(a_y/a_x)^2$.

The magnetization dynamics is described by the phenomenological Landau-Lifshitz-Gilbert equation in angles $\theta $ and $\varphi $ representation ($m_{x}+\rmi m_{y}=\rme^{\rmi\varphi}\cos\theta$ and $m_{z}=\sin \theta$):
\begin{equation}
\label{LLG}
\eqalign{
        \dot{\theta}\cos\theta=\gamma_e \partial_{\varphi}W/M_0+\alpha_{\rm G}\dot{\varphi}\cos^2\theta \cr
        \dot{\varphi}\cos\theta=-\gamma_e \partial_{\theta}W/M_0-\alpha_{\rm G}\dot{\theta},
        }
\end{equation}
where $\gamma_e=2\mu_B/\hbar$ is the electron gyromagnetic ratio and $\alpha_{\rm G}\approx0.01$ is the dimensionless Gilbert damping constant \cite{oogane2006,nibarger2003}.

The typical value of the out-of-plane component of magnetization unit
vector is $m_{z}^2 \sim \omega _{\mathrm{ex}}/\omega _{A}\approx 0.01$, which is estimated by comparing two terms in \eref{energy}.
Since $|m_z|\ll1$, one could make an appropriate linearization: $m_{x}+\rmi m_{y}\approx \rme^{\rmi\varphi}$ and $m_{z}\approx \theta$.
Using \eref{energyE} we rewrite the equations~\eref{LLG}:
\begin{equation}
\label{systemf}
\eqalign{
        \dot{\varphi}=-\omega_A\theta\left(1-K\sin^2\varphi-\frac{\omega_{\mathrm{ex}}E}{\omega_{A}E_0}\sin(\varphi-\beta)\right)-
        \alpha_{\rm G}\dot{\theta} \cr
        \dot{\theta}=\omega_{\mathrm{ex}}\frac{E}{E_{0}}\cos(\varphi-\beta)+\omega_A\frac{K}{2}\sin2\varphi+\alpha_{\rm G}\dot{\varphi},
        }
\end{equation}
where $\omega _{A}=4\pi \gamma M_{0} \approx 2\pi \times 28$~GHz and ${\omega _{\mathrm{ex}}=J_{\mathrm{ex}}n_{\mathrm{el}}s_0/\hbar \approx 2\pi \times 0.35}$~GHz are the effective anisotropy and exchange frequencies respectively and $\left( \varphi -\beta \right) $ is the angle between $\bi{m}$ and $\bi{E}$ as shown in the figure~\ref{Schema}.
Therefore in-plane rotational dynamics of magnetization is determined mainly by the out-of-plane deviation $\theta $, dynamics of which is controlled by the spin polarized current as can be seen in Eqs.~\eref{systemf}.

Considering $\omega _{A}\gg\omega _{\mathrm{ex}}$ and $\alpha_{\rm G},K\ll1$, the equations~\eref{systemf} can be reduced to the Hamilton-like form with dissipation:
\begin{equation}
\label{system}
\eqalign{
        \dot{\varphi}=-\omega _{A}\theta \cr
        \dot{\theta}=\omega _{\mathrm{ex}}\frac{E}{E_{0}}\cos \left( \varphi -\beta \right)+\omega_A\frac{K}{2}\sin2\varphi +
        \alpha_{\rm G}\dot{\varphi}.
        }
\end{equation}
Here $\varphi $ and $-\theta $ play the roles of the generalized coordinate and momentum, correspondingly, then $\omega _{A}^{-1}$ is the analog of the inertial mass and $\omega_{\rm{ex}}E(\tau)/E_{0}$ is the amplitude of driving force.

Resulting dynamic equation for the in-plane rotation of magnetization angle $\varphi $ derived from Eqs.~\eref{system} is:
\begin{equation}
    \ddot{\varphi}+\alpha_{\rm G}\omega _{A}\dot{\varphi}+\omega^{2}_{A}\frac{K}{2}\sin2\varphi+
    \omega _{A}\omega _{\mathrm{ex}}\frac{E}{E_{0}}\cos \left(\varphi -\beta \right)=0.
    \label{pend}
\end{equation}
This result is consistent with the approach of Bazaliy \cite{bazaliy2007,bazaliy2011,bazaliy2012}, where it was shown that planar spin transfer devices with dominating easy-plane anisotropy can be described by an effective one-dimensional equation for the in-plane magnetization.

The last term in \eref{pend} is a driving rotation force that is proportional to the ratio of the drift increment of the Bychkov-Rashba SOI energy density in the electron subsystem to the anisotropy energy density in the magnetic subsystem.
It also indicates that the electric field excites the in-plane rotation of the magnetization by transferring the increment Rashba SOI via exchange interaction to the energy of out-of-plane magnetization deviation and this effect is suppressed by the damping.
Therefore the magnetization dynamics is a superposition of in-plane precession and out-of-plane nutation.

We introduce dimensionless time $\tau =\alpha _{\rm G}\omega_{A}t/2$ then \eref{pend} is transformed to:
\begin{equation}
\ddot{\varphi}+2\dot{\varphi}+\Delta(\tau)\cos(\varphi-\beta)+\chi\sin2\varphi=0, \label{dless}
\end{equation}
where $\Delta(\tau)=4\omega_{\mathrm{ex}}E(\tau)/\alpha_{\rm G}^{2}\omega_{A}E_0$ and $\chi=2K/\alpha_{\rm G}^{2}$. The value of dimensionless coefficient $\Delta (\tau )$ in \eref{dless} according to the typical heterostructure parameters provided above is $\Delta (\tau )\approx 10^3 \cdot E/E_{0}$.
It is limited due to the condition imposed above that drift increment of electron momentum should be much smaller than the thermal one (i.e. $E\ll10$~kV/cm).
The value of coefficient  $\chi$ resulting from shape anisotropy is $\chi(\eta)\approx600\eta^2(1+0.75\eta^2)$ where $\eta$ is an eccentricity according to \eref{energyE}. It is relatively high even for the element in-plane shape close to circular.

The equation \eref{dless} possesses the following symmetry: inverse in-plane magnetization rotation ($\varphi\rightarrow-\varphi$) is caused by applying the electrical field at an angle $\beta\rightarrow\pi-\beta$. Generally one can change the direction angle $\beta$ of electric field $\bi{E}$ by applying different alternating voltages to two perpendicular pairs of contact microstrips embedded in the hybrid structure.
This can be treated as an additional possibility of magnetization switching induced by the electric field rotation, but in this work we focus on the case of constant $\beta$ for the sake of considered model simplicity.

If the ferromagnetic particle has circular in-plane shape ($\chi=0$) then its magnetic ground state is infinitely degenerate: the system does not have preferential in-plane direction.  For constant $\beta$ we can replace $\varphi \rightarrow {\xi =\varphi -\beta +\pi /2}$ thus \eref{dless} is transformed to:
\begin{equation}
    \ddot{\xi}+2\dot{\xi}+\Delta (\tau )\sin \xi =0.
\label{dimensionless}
\end{equation}
This equation describes the classical parametrically forced pendulum.
We take initial conditions $\xi_{0}\equiv \xi (0)=\pi /2-\beta $ and $\dot{\xi}(0)=0$.

\begin{figure}[ht]
\center
\includegraphics[width=8 cm]{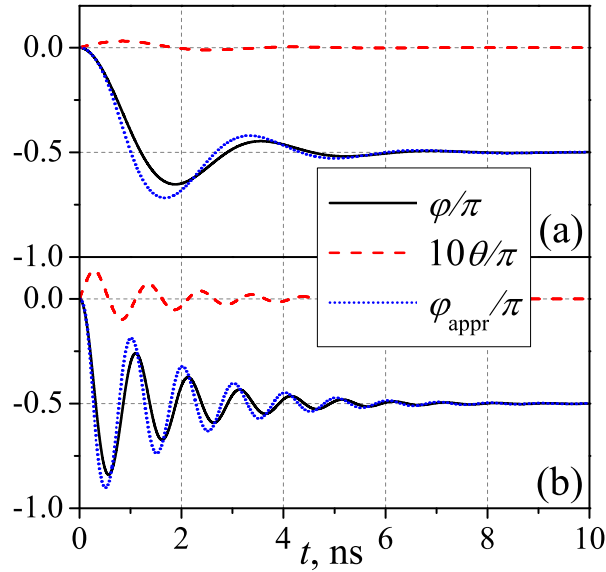}
\caption{Time dependence of magnetization direction angles in the sufficiently large applied fields:
\textbf{(a)} ${E=70}$~V/cm ${(\Delta_0=10)}$
\textbf{(b)} $E=0.7$~kV/cm ($\Delta_0=100$). Angle $\varphi_{\mathrm{appr}}$ is the approximate solution given by \eref{approx}.}\label{Elarge}
\end{figure}

For constant electric field $\Delta (\tau )=\Delta _{0}>0$ one can use approximate solution of linearized \eref{dimensionless}:
\begin{equation}
\label{approx}
\tan\frac{\xi}{2}\approx e^{-\tau }\left( \cos\Omega \tau +\frac{\sin \Omega \tau }{\Omega }\right) \tan \frac{\xi _{0}}{2},
\end{equation}
where $\Omega=(\Delta _{0}-1)^{1/2}$ is a frequency of damped oscillations shown in the figure~\ref{Elarge}.

\begin{figure}[ht]
\center
\includegraphics[width=8 cm]{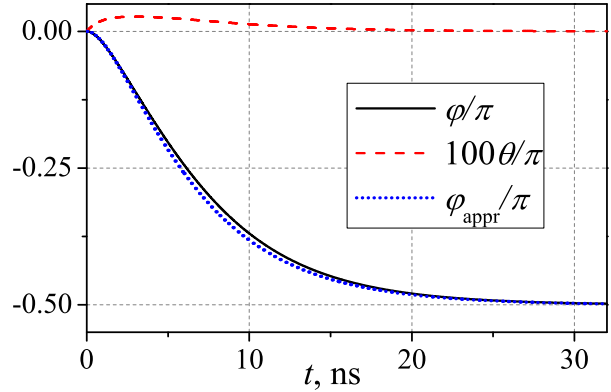}
\caption{Time dependence of magnetization direction angles in a sufficiently small applied field $E=3.5$~V/cm ($\Delta_0=0.5$). Angle $\varphi_{\mathrm{appr}}$ is the approximate solution by \eref{approx}.}
\label{Elow}
\end{figure}

Expression \eref{approx} is also the approximate solution of \eref{dimensionless} if $\Delta _{0}<1$.
In this case $\Omega $ is the imaginary and trigonometric functions are replaced with hyperbolic, resulting in the aperiodic damping shown in the figure~\ref{Elow}.

If ${\Delta _{0}>1}$, i.e. the applied electric field magnitude is high enough ($E>7.5$~V/cm), magnetization rotates to the maximal
angle $|\Delta \varphi _{\max }|\approx \xi _{0}+2\arctan \left( e^{-\pi/\Omega }\tan \xi _{0}/2\right)$ during the time $\tau _{\mathrm{rot}}=\pi/\Omega $ which is about
${t_{\mathrm{rot}}\approx \pi \left( \omega_{\mathrm{ex}}\omega _{A}E/E_{0}\right) ^{-1/2}}$.
Since the magnetization rotation is inertial according to \eref{dimensionless}, maximal rotation angle could be larger than $|\Delta \varphi _{\max }|$ if the applying electric field pulse time $\tau_{\mathrm{pulse}}<1$ so $\dot{\varphi}$ is not negligible.
For the aperiodic damping ($E<7.5$~V/cm), the charac\-teri\-stic time of magnetization rotation at $\pi/2$ to the electric field direction is $\tau _{\mathrm{damp}}=4/\Delta _{0}$ which is about ${t_{\mathrm{damp}}\approx 2\alpha_{\rm G}\left(\omega_{\mathrm{ex}}E/E_{0}\right) ^{-1}}$ in the physical units.

The difference between these characteristic times is that $t_{\mathrm{damp}}$ depends on the Gilbert damping $\alpha _{\rm G}$ and does not depend on magnetic anisotropy $\omega _{A}$ unlike $t_{\mathrm{rot}}$.
This effect can be explained by an analogy with a classical oscillator.
The rotation time $t_{\mathrm{rot}}$ is the first turning point time i.e. half of period of the oscillations. It depends on $\omega _{A}$ which is the inertia coefficient (reciprocal mass) unlike the decay time $t_{\mathrm{damp}}$ determined only by the ratio of the damping to the restoring force coefficients.

\begin{figure}[ht]
\center
\includegraphics[width=8 cm]{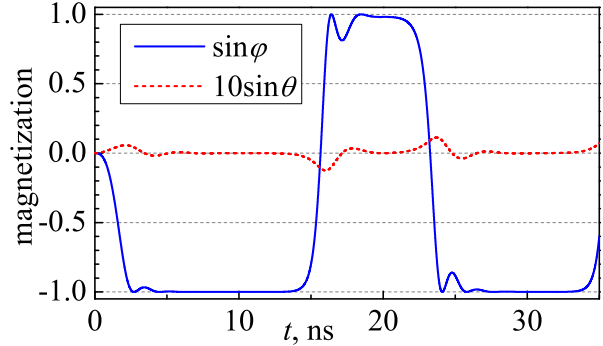}
\caption{Time dependence of normalized magnetization components $\sin\varphi=M_y/M_0$ and $\sin\theta=M_z/M_0$ in the
alternating applied field $E(t)=E_{\mathrm{amp}}\sin (\omega t)$, $E_{\mathrm{amp}}=70$~V/cm, $\omega = 2\pi\times0.05$ GHz.}
\label{Ealt}
\end{figure}

Interesting possibilities of the magnetization controlling and parametric resonance appear for strong enough alternating electric fields (${E\gg 10}$~V/cm) \cite{broer2004,mclaughlin1981}.
According to \eref{pend} the equilibrium direction of the magnetization in the electric field is ${\varphi_{\mathrm{eq}}=\beta-\pi /2}$, so it reverses after the electric field reversal during the time of the order of magnitude of $t_{\mathrm{damp}}$.
Therefore in the alternating electric fields with high enough amplitude and low frequency $\omega_{E}$ the time profile of magnetization takes the form of rectangular pulses.
We assume that periodic alternating function $\Delta(\tau)\propto E(\tau)$ is a slowly varying function on a short time interval $[\tau,\tau+\delta\tau]$ $\left(\delta\tau\ll T_E\right)$ and $\xi\ll1$ ($\sin \xi \approx \xi$).
Then local solution of \eref{dimensionless} at this interval is:
\begin{equation}
    \xi(\tau+\delta\tau)\sim\xi(\tau)\exp\left[-\left(1\pm\sqrt{1-\Delta(\tau)}\right)\delta\tau\right].
\end{equation}
The requirement of damped motion is the positive logarithmic decrement:
\begin{equation}\label{decrement}
    \ln \left|\frac{\xi(\tau)}{\xi(\tau+T_E)}\right|\sim\int\limits_{0}^{T_E}\mathrm{Re}\left[1-\sqrt{1-\Delta(\tau)}\right]d\tau>0.
\end{equation}
The estimation \eref{decrement} gives the boundary values for electric field amplitude when solution \eref{dimensionless} is still damped.
It gives $E_{\rm{amp}}<20$~V/cm ($\Delta_0<3$) for the case of periodic rectangular electric field pulses $\Delta(\tau)=\Delta_0\mathrm{sgn}(\sin\omega_E\tau)$ and $E_{\rm{amp}}<27$~V/cm ($\Delta_0<4$) for the harmonic applying field $\Delta(\tau)=\Delta_0\sin\omega_E\tau$.
In higher fields magnetization switches between two opposite directions perpendicular to $\bi{E}$ and its time dependence has rectangular pulse form as it is shown in the figure~\ref{Ealt}.

The estimate \eref{decrement} properly describes only the case of low frequencies $\omega_E \ll 2\pi \times 1$ GHz.
Another interesting behavior can be seen for sufficiently large frequencies $\omega_{E}^{2}\sim\Delta_{0}$.
One can apply the superposition of direct and alternating electric fields and obtain a realization of the well-known classical parametrically forced pendulum with rich dynamical behavior \cite{broer2004,mclaughlin1981}.

We will consider further that the particle has elliptical in-plane shape ($\chi\neq0$) and electric field $\bi{E}$ is a constant vector after switching on. Therefore the ground state of magnetization is double degenerate. In the case of small deviations of the in-plane angle $\varphi$ we obtain solution similar to \eref{approx}:
\begin{equation}\label{appr_chi}
    \varphi_{\mathrm{appr}}=\varphi_{\mathrm{inf}}\left[1-\left(\cos\Omega\tau+\frac{\sin\Omega\tau}{\Omega}\right)e^{-\tau}\right],
\end{equation}
where $\Omega^2=\Delta\sin\beta+2\chi-1$ and $\varphi_{\mathrm{inf}}=-\Delta\cos\beta/{(\Omega^2+1)}$. There are few significant differences between results \eref{appr_chi} and \eref{approx}: the frequency $\Omega$ in \eref{appr_chi} depends not only on the magnitude of the electric field $E$, but also on its direction $\beta$, and the value of this $\Omega$ is higher due to the relatively high shape anisotropy $\chi$.

The solution \eref{appr_chi} is applicable when $\varphi_{\mathrm{inf}}\ll1$ i.e. $\Delta/\chi\ll1$ or $\beta$ close to $\pi/2$.
In order to examine an interesting case of switching between the two ground states of magnetization, instead of direct solving \eref{dless} we describe the dynamics by using the mechanical analogy \cite{bazaliy2007}.
One can obtain after integration of \eref{dless} with initial conditions $\varphi(0)=\dot{\varphi}(0)=0$:
\begin{equation}\label{int_dless}
    \frac{\dot{\varphi}^2}{2}+2\int\limits_{0}^{\tau}\dot{\varphi}^2d\tau+V(\varphi)=0,
\end{equation}
where $V(\varphi)=\Delta(\sin(\varphi-\beta)+\sin\beta)+\chi\sin^2\varphi$. The first term in the \eref{int_dless} is a kinetic energy, the second one is an integral of the Rayleigh dissipation function over time and the third one is a potential energy by analogy with classical mechanics. Thus the dynamics of $\varphi$ is possible if $V(\varphi)<0$ since the first two terms in \eref{int_dless} are positive. The profile of potential energy $V(\varphi)$ depends on the two changeable parameters $\Delta$ and $\beta$ -- this defines dynamics diversity.
\begin{figure}[ht]
\center
\includegraphics[width=8 cm]{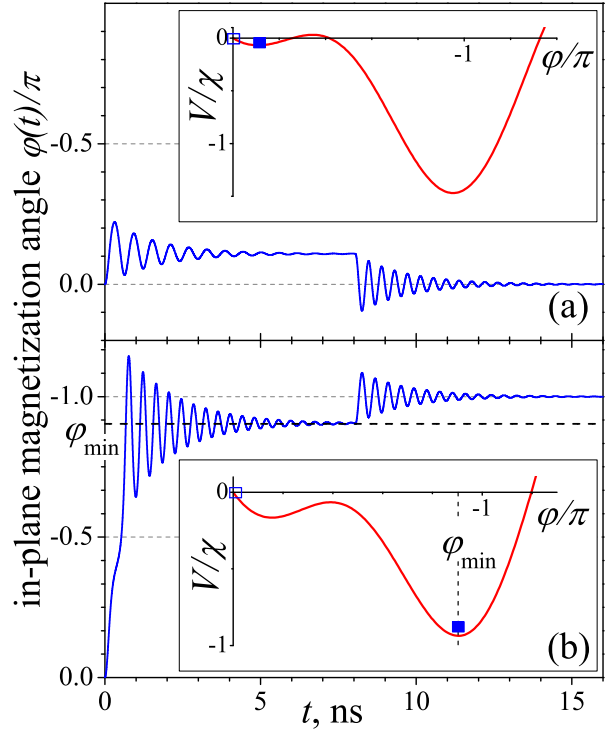}
\caption{Dependence of magnetization angle $\varphi$ on time $t$. The particle in-plane shape is an ellipse ($a_y/a_x=0.8$). The electric field $E=1.7$~kV/cm ($\Delta/\chi=0.83$) was applied during $t_E=8$~ns at the angle (a) $\beta=-\pi/3$ and (b) $\beta=-\pi/6$ to the initial direction of magnetization. The potentials $V(\varphi)$ (introduced in \eref{int_dless}) are represented in the corresponding insets.}
\label{Switchex}
\end{figure}

Let us examine the possibility of magnetization switching between two ground states ($\varphi=0$ and $\varphi=\pi$) of the elliptically elongated magnetic particle.
The analysis of the potential $V(\varphi)$ shows that $V(\varphi)$ has two local minima if $\Delta/\chi<2$.
In the vicinity of $\beta=0$ if $\Delta/\chi>1/(1-\beta)$ it allows transfer from the initial state ($\varphi=0$) to the second minimum closest to $\pi$.
Then, after switching off electric field, the in-plane magnetization will turn to $\varphi=\pi$.
This case is shown in the figure~\ref{Switchex}(b). Otherwise, if both local maxima of $V(\varphi)$ are positive, the switching does not occur, as it is shown in the figure~\ref{Switchex}(a).

\begin{figure}[ht]
\center
\includegraphics[width=8 cm]{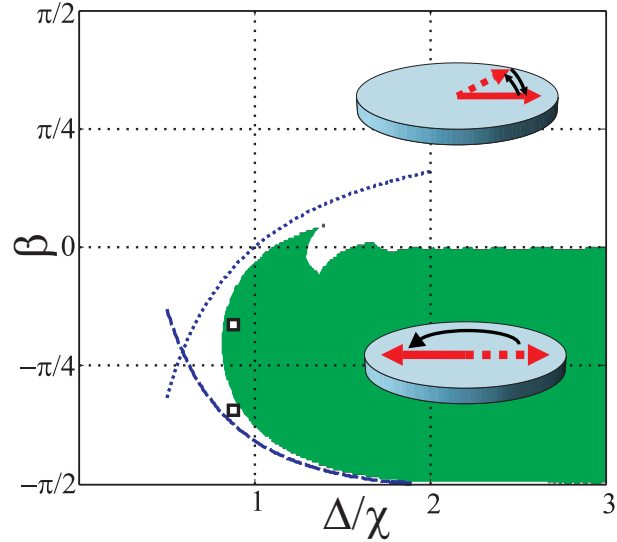}
\caption{Switching diagram for the parameters $\Delta/\chi$ and $\beta$. The parameter $\Delta/\chi$ is proportional to the applied field $E$ and $\beta$ is the angle between $\bi{E}$ and $\bi{m}_0$. White region corresponds to returning into the initial position and filled (green) region corresponds to 180 degrees switching of magnetization. Dotted and dashed lines show the parametric boundaries obtained in asymptotic expansion near $\beta=0$ and $\beta=-\pi/2$, respectively. Square markers correspond to the applied fields which initiate magnetization dynamics shown in the figure~\ref{Switchex}.}
\label{ParamDiag}
\end{figure}

The parametric diagram shown in the figure~\ref{ParamDiag} depicts the final in-plane magnetization state in the elliptical particle after the electric field pulse.
Duration of the applied field pulse is 8~ns that is longer than relaxation time
$2/(\alpha_{\rm G}\omega_A)\approx1.6$~ns and sufficient for the rotation of magnetization into new equilibrium position.
The diagram coordinates $\Delta/\chi=2\omega_{\mathrm{ex}}E/\omega_A K(\eta)E_0$ and $\beta$ represent magnitude and direction of applied electric field.
The choice of $[-\pi/2,\pi/2]$ interval for plotting $\beta$ is due to the symmetry of \eref{dless}.
The minimal electric field $E_{\mathrm{min}}$ required for the switching corresponds to the $\Delta/\chi=0.81$ as it shown in the figure~\ref{ParamDiag}. Considering the expansion of $K(\eta)$ given in \eref{energyE}, the estimation for this minimal field is ${E_{\mathrm{min}}\approx4\eta^2}$~kV/cm for in-plane particle shape close to circle (the eccentricity $\eta\ll1$).

In summary, we have studied the current-induced magnetic dynamics of ferromagnet/doped semiconductor multilayer structure within a practical model and obtained the scaling relations for the magnetization dynamics in terms of the system parameters.
We have shown that the magnetization rotates to the angle orthogonal to the electric field direction during the characteristic time $t_{\mathrm{damp}}\approx 2\alpha_{\rm G}\left(\omega _{\mathrm{ex}}E/E_{0}\right)^{-1}\approx E^{-1}\times45$~ns$\cdot$V/cm.
The time profile of magnetization takes the "balanced" rectangular waveform for the low-frequency alternating electric field.
The magnetic moment of the particle with elliptical in-plane shape with small eccentricity $\eta$ can be switched to the opposite direction by the electrical field pulse applied at a negative angle to the initial direction of magnetic moment with duration $\sim10$~ns and magnitude $E>4\eta^2$~kV/cm.
\ack
E.Y.S. acknowledges support of the University of the Basque Country UPV/EHU under program UFI 11/55, Spanish MEC/FEDER (FIS2015-67161-P) and Grupos Consolidados UPV/EHU del Gobierno Vasco (IT-472-10). P.V.B. acknowledges support from the Erasmus Mundus Action 2 ACTIVE project.
\section*{References}
\bibliography{spintron2}
\end{document}